\documentclass[pre,twocolumn,showpacs]{revtex4}
\let\a=\alpha \let\g=\gamma \let\d=\delta
\let\e=\varepsilon \let\z=\zeta \let\h=\eta
\let\th=\vartheta\let\k=\kappa \let\l=\lambda \let\m=\mu \let\n=\nu
\let\x=\xi \let\p=\pi \let\r=\rho \let\s=\sigma \let\t=\tau
   
  \let\D=\Delta \let\Th=\Theta
\let\L=\Lambda  \let\Si=\Sigma \let\F=\Phi
 \let\O=\Omega 
\let\dpr=\partial\let\0=\noindent\let\fra=\frac
\def\V#1{{\,\underline#1\,}}\def\lis#1{\overline#1}

\def\*{\vskip3mm}
\def\Onlinecite#1{[\onlinecite{#1}]}

\newcommand\eg{{\it e.g.\ }}\newcommand\ie{{\it i.e.\ }}
\newcommand\dQ{{\not \kern-1.5pt d\, Q}}
\newcommand\dq{{\not \kern0.pt d\,Q}}
\newcommand\defi{\,{\buildrel def \over =}\,}
\newcommand\revtex{{R\kern-1mm\lower0.5mm\hbox{E}\kern-0.6mm V\kern-0.5mm%
\lower0.5mm\hbox{T}\kern-0.5mm E\kern-.5mm \lower0.5mm\hbox{X}}}

\newdimen\xshift \newdimen\xwidth \newdimen\yshift
\def\ins#1#2#3{\vbox to0pt{\kern-#2 \hbox{\kern#1
#3}\vss}\nointerlineskip}
\def\eqfig#1#2#3#4#5{ \par\xwidth=#1
\xshift=\hsize \advance\xshift by-\xwidth \divide\xshift by 2
\yshift=#2 \divide\yshift by 2 \hbox{\hglue\xshift \vbox to #2{\vfil #3
\includegraphics{#4.ps} }\hfill\raise\yshift\hbox{#5}}}
\def\8{\write13}

\def\NN{{\cal N}}\def\PP{{\cal P}}
\begin{document}\relax 

\preprint{RU/3-03g} 

\title{Entropy production in nonequilibrium thermodynamics: a review}

\author{Giovanni Gallavotti} 
\affiliation{Rutgers Hill Center, I.N.F.N. Roma1, Fisica Roma1}

\date{\today}
\begin{abstract}
Entropy might be a not well defined concept if the system can undergo
transformations involving stationary nonequilibria. It might be
analogous to the heat content (once called ``caloric'') in
transformations that are not isochoric (\ie which involve mechanical
work): it could be just a quantity that can be transferred or created,
like heat in equilibrium. The text first reviews the philosophy behind
a recently proposed definition of entropy production in nonequilibrium
stationary systems. A detailed technical attempt at defining the
entropy of a stationary states via their variational properties
follows: the unsatisfactory aspects of the results add arguments in
favor of the nonexistence of a function of state to be identified with
entropy; at the same time new aspects and properties of the phase
space contraction emerge.
\end{abstract}


\maketitle
\def\CHAOS#1{}
\CHAOS{The problem of defining ``Entropy'' in systems out of
equilibrium but in stationary states has not been solved: indeed no
universally agreed upon way of defining this notion seems to be
available. If entropy has to be related to heat tranfer as in
equilibrium Thermodynamics then entropy of a nonequilibrium stationary
state which produces heat at a steady rate might just be not
definable.  Here I consider mainly the problem for systems that are
subject to external non conservative positional forces and to other
mechanical forces which take out the energy provided by the work done
by external forces, thereby constituting a mechanical model of a
``thermostat''. Such systems have been extensively studied in recent
times in the literature on computational Physics. To study their
stationary states and the transformations which bring from one
stationary state to another, through intermediate stationary states
(``quasi static transformations''), I adopt a precise definition of
``entropy production rate'' proposed and employed also in the
mentioned literature. It is argued that this definition is compatible
with the ordinary definition of entropy for the familiar,
phenomenologically defined, thermostats when only equilibrium states
(and quasi static transitions, through intermediate equilibria,
between them) are considered; but it leads to attribute to a general
stationary nonequilibrium state an entropy $-\infty$: hence it does
not allow defining a nontrivial entropy of the state. This does not
contradict the possible existence of $H$--functions, \ie of functions
of the microscopic state of the system which, when the system is
started in a given configuration, approach a maximum value
monotonically, apart from rare fluctuations; and the maximum is
attained when the system microscopic state reaches configurations
typical of the stationary distribution which pertains to the
parameters characterizing the stationary state. In a technical section
I provide an example of $H$--function which requires setting a precise
definition of ``coarse graining'': and discuss why the example falls
short of deserving being called an entropy function, although it is a
good $H$--function and it is also related to the above entropy
production rate. The conclusion is that in stationary nonequilibria
and transformations between them entropy may be undefined while
entropy production rate is well defined. More general external forces
must be considered if one wants to deal with heat conduction problems:
an example is also briefly discussed.  The status of entropy in
nonequilibrium thermodynamics seems similar to the status of heat in
equilibrium thermodynamics: one can define heat (respectively entropy)
content only if the transformations considered are isochoric, \ie
through states that have the same volume (respectively through states
of thermodynamic equilibrium) while in more general transformations
only heat transfer (respectively entropy transfer) is well defined. A
technical comparison with other approaches is attempted.  \*\* }

\centerline{\bf I. Thermostats and chaotic hypothesis.}
\*
In studying equilibrium and nonequilibrium thermodynamics the notion
of {\it thermostat} plays an important role: it is usually defined
empirically \Onlinecite{Ze68} as a physical system capable only of
exchanging heat without changing temperature or performing work (hence
it is ideally an infinite system). 

However one can also envisage concrete mechanisms to keep a system in
a stationary state, realized by a mechanical force of nonconservative
nature.  Here I want to consider mechanical systems which, in spite of
being acted upon by nonconservative forces, are kept in a stationary
state by other mechanical forces, and to study which relations, if
any, can be established between the various stationary states,
\Onlinecite{Ga03}. The transformations between the stationary states
will be ``{\it quasi static}'' trasformations through intermediate
stationary states.

I shall consider only systems consisting of many particles and I shall
{\it not} consider systems that are modeled by continua: because the
latter are more complex and can be regarded as an idealization in
which several many particles systems are put together, {\it one per
``volume element''}, and studied on time and space scales vastly
different from the ones we consider for the evolution of the simple
systems which we imagine to constitute the volume elements of the
continuum, \Onlinecite{Ga02}. Continua can also be considered but one
first must understand the thermodynamics of a simple system,
\Onlinecite{Ga02}.

A simple system will be described by a differential equation in its
phase space: we write it as $\dot x=X_E(x)$ where $x=(\dot{\V q},\V
q)\in R^{6N} \equiv \O$ ({\it phase space}), $X_E(x)=(\V Y_E(x),\V
X_E(x))$, $N$=number of particles, mass of the particles $m=1$, with

$$\ddot \V q=\V f(\V q)+\V E\cdot \V g(\V q)-\th_{\V E}
(\dot {\V q},\V q)= \V X_E(x)$$
where $\V f(\V q)$ describes the internal (conservative) forces (\eg hard
cores), $\V E\cdot\V g(\V q)$ represents the ``external forces''
(nonconservative) acting on the system, with $\V E$ being their
``strength'': for definiteness we suppose that they are locally
conservative (like an electromotive force) but not globally such, and
$\V\th_{\V E}$ is the force law which models the action exerted by the
thermostat on the system to keep it from indefinitely acquiring
energy: this is why I shall call it a {\it mechanical
thermostat}. Linearity in $\V E$ does not mean that I am assuming the
forces to be small (the theory of linear nonequilibrium is amply
discussed in the literature, \Onlinecite{DGM84}): in any event the
response curves of forces versus fluxes will be non linear evenfor the
abobe model. 

More generally the external forces could be velocity dependent and
even periodically time dependent: we first restrict to positional
forces for simplicity. Velocity dependent forces should also be
considered as they are necessary to study heat conduction problems:
this will be briefly discussed in Sec. VI.  \*

{\it Assumption {\rm (chaotic hypothesis, \Onlinecite{GC95})}: The
system evolution is assumed to be ``as chaotic as possible'', \ie it is
assumed to be hyperbolic.}  \*

Technically I mean that the system is supposed to be ``a transitive
Anosov system'', \Onlinecite{Ga00,Ga02}.  

We shall denote temperature by $\Th$ to avoid confusion with time
which will be often denoted $T$. Models of thermostats in the above
sense can be very different even for the same macroscopic system; for
instance (a list far from exhaustive)

(1) assuming the system to have hard cores one can suppose that the
   collisions are inelastic: the head-on component of the energy is
   decreased by a scale factor $\h<1$ upon each collision or, {\it
   alternatively}, the total energy of the two colliding particles is
   rescaled and assigned a given value $3 k_B \Th$. This kind of
   thermostat has been first introduced and applied in ``Drude's
   electrical condution model'' to model the thermostatting effects of
   the collisions electron--phonon by Drude and, later, by Lorentz,
   \Onlinecite{Be64}; or

(2) assuming that there is a background friction $\th_i=-\n \dot
    q_i$, $\n>0$, for all components of $\ddot q_j$; or

(3) assuming {\it least effort} to keep, say, the total kinetic
    energy or the total energy constant, (``{\it Gaussian
    thermostat}'', \Onlinecite{EM90}).

Stochastic thermostats could also be allowed (they just add degrees of
freedom to the equations) but we do not consider them here,
\Onlinecite{Ku98,LS99,Ma99,De02}.
\*
\centerline{\bf II. SRB statistics and nonequilibrium ensembles}
\*
Any initial state $x$, randomly chosen in phase space with a
probability distribution which has a density in phase space (in jargon
``absolutely continuous distribution with respect to Liouville
measure''), {\it will admit a statistics} (under the above chaotic
assumption): \ie for all (smooth) observables $F$

\begin{eqnarray}
\lim_{T\to\infty} \frac1T\int_0^T F(S_tx) dt=\int_\O \m_{\V E}(dy) F(y)
\label{a1}\end{eqnarray}
where $\m_{\V E}$ is a stationary probability distribution on phase
space, called the {\it SRB distribution} or {\it SRB
statistics}, \Onlinecite{Ru95,Ga00,Ga02,GBG04} .
\* 

{\it Definition: A system in a microscopic state $x$ which has SRB
  statistics $\m_{\V E}$ is said to be in the stationary state $\m_{\V
  E}$. The collection of all stationary states of a system that are
  constructed by varying the parameters (typically the volume $V$ of
  the container, the particles number $N$, the external forces $\V E$,
  {\it etc}) will be called a ``nonequilibrium ensemble''.}
\* 

\0{\it Remark:} Hence here an {\it ensemble} is a collection of
probability distributions {\it although}, often, established
terminology indicates an ensemble to be {\it a single element} of the
collection: according to the above definition we would simply call, in
equilibrium cases, ``microcanonical ensemble'' what with the usual
teminology is called ``the collection of the microcanonical
ensembles as the parameters $V,U$ vary''. The latter circumvolution
being in my view too awkward I have adopted calling an ensemble
already the whole collection of distributions (as the French word
ensemble literally suggests).  \*

The notion of nonequilibrium ensemble is wider than in equilibrium as
it depends {\it also} on the equations of motion, hence on the
thermostat model. Therefore one expects that, as it happens in
equilibrium statistical mechanics, there should be ``equivalent
ensembles'' corresponding to classes of different meaningful models
for thermostats acting on a system, see \Onlinecite{Ga00,Ga02} and
appended references.

Equilibrium is a special case of nonequilibrium: in such case $\V E=\V
0$ and $\V\th_{\V E}=\V0$ and the chaotic hypothesis implies the
validity of the ergodic hypothesis and the family of SRB distributions
can be parameterized by total energy $U$ and volume $V$ and
coincides with the microcanonical ensemble, \Onlinecite{Ga00}.

We now want to consider which relations can be established in general
between the properties of stationary states that can be transformed into
one another by changing the external parameters.

If we limit ourselves to equilibrium states then it is well known
since Boltzmann (in his papers in the period 1866--1884, see
\Onlinecite{Bo84}) that if a transformation generates an energy
variation $dU$ and a volume variation $dV$ when the pressure (defined
microscopically) is $p$ and the average kinetic energy is $\frac32 N
k_B\Th$ then, \Onlinecite{Ga00},

$$\fra{dU+p\, dV}{\Th}={\rm exact}$$
while $dU + p\, dV$ is not exact, {\it except} in the isochoric case
(\ie when $dV=0$), and it is called the {\it heat transferred} from the
reservoirs to the system. It makes no sense to talk of amount of heat
contained in the system unless one limits oneself to studying isochoric
transformations: there is no {\it caloric} (\ie no {\it heat content},
see \Onlinecite{Br03,Gr798})
unless one allows only the latter type of transformations in which the
system performs no work (and in that case it is just another name for
internal energy).

Defining {\it entropy} as a primitive of the exact differential $(dU+p\,
dV)/\Th$, the first immediate question is whether one can extend the notion of
{\it entropy} content to non equilibrium states.
\*

\centerline{\bf III. Entropy production and temperature.}
\*

The proposal that emerges from recent literature (mostly based on
numerical experiments), \Onlinecite{An82,EM90,Ru99,Ga00}, is to
define, if $k_B$ is Boltzmann's constant, \*

{\it Definition: The entropy production rate $s$ in a stationary state
  $\m_{\V E}$ is $s=k_B\s$ with

$$\s=\int_\O \m_{\V E}(dx)\,\s(x)$$ 
where $\s(x)=-$ divergence of $X_{\V E}(x)$ and $\m_{\V E}$ is the SRB
statistics.}  \*

Note that $\s$ is also the average value over time of the phase space
contraction, by (\ref{a1}).

An important general theorem,\Onlinecite{Ru96}, guarantees
that $\s\ge0$, and $\s=0$ corresponds to the case in which the SRB
distribution $\m_{\V E}$ admits a density on phase space, a case that
one naturally identifies with an equilibrium state and which
essentially happens only if $\V E=\V0$.

The above definition leads to a natural definition of temperature of
the thermostatting forces, \Onlinecite{FN03,Ga03}: note that there is
no universally accepted definition of temperature in systems out of
equilibrium, even if stationary. Namely one sets \*

{\it Definition: the effective temperature $\Th$ of the thermostats
equilibrating the external forces keeping a system in a stationary
nonequilibrium state is

$$ \Th=\frac{W}{s}$$
where $W$ is the work per unit time done on the system by the external
forces and $s$ is the entropy production rate.}
\*

A connection between temperature and entropy production has been
already hinted and even used in the literature (\Onlinecite{CKP97,
FN03} and F. Bonetto) and alternative definitions have been proposed,
see \Onlinecite{Cu02}.

The above definition does not make sense as such in equilibrium
because it becomes $0/0$: however one can imagine introducing a small
forcing and a corresponding thermostat. Then in the limit of vanishing
forcing this yields a definition of $\Th$ which by the ``fluctuation
dissipation theorem'' would be correct,
\Onlinecite{CKP97,Ga96,Ga96a,GR97,ZRA04}.

A natural question is whether the above definition of temperature is 
related to heat transfer: does a heat current  arise when two systems
at different temperature in the above sense are thermaly connected
(\ie are made to interact)? The question is difficult and an example
in which the answer is affirmative in discussed in Section VI, along
the lines of \Onlinecite{Ga96}.

Adopting the above concepts leads naturally to giving up the
possibility of defining entropy content of a system: in nonequilibrium
thermodynamics entropy ends up to be undefined and one can speak only
of ``entropy production'' or ``transfer'': much as ``caloric'' or ``heat
content'' is undefined in equilibrium thermodynamics. Should one
insist in defining the entropy content of a dissipating (\ie with
$\s>0$) stationary state one would be compelled to assign to it a
value $-\infty$, because the system cedes entropy at a constant
rate.

\*

\centerline{\bf IV. $H$-functions.}
\*

The above is not in contradiction with the possibility of existence of
a function which, given an initial state $x$, will evolve
monotonically until reaching a maximum value, the same for almost all
$x$ in phase space, \Onlinecite{Le93, GL03, GGL03}: and this is not in
contrast with microscopic reversibility. For instance in the case of
the evolution of a rarefied gas we can imagine to divide the one
particle phase space into cells $C$, ``of appropriately chosen size $|C|$'',
\Onlinecite{Le93}, and call $f_C(x)$ the occupation number of each
cell by the particles in $x$. Then, if $t\to S_t x$ 
denotes the time evolution of the initial data $x$ at time $t$,

$$-k_B \fra1t \int_0^t\sum_C f_C(S_\t x) \log f_C(S_\t x)\, dt$$
will converge, and for practical purposes monotonically after a short
transient, as $t\to\infty$ to a limit which, if the Boltzmann equation
holds within a good approximation, maximizes $-k_B \sum_C p_C \log
p_C$, $p_C\ge0$, (subject to the conservation constraints like $\sum_C
p_C=N$, {\it etc.}, \Onlinecite{Le93}) and the limit value is given
by the entropy $S$ of the equilibrium state associated with $x$. If
Boltzmann's equation is (unreasonably) dismissed then still the above
quantity will converge to essentially the same limit but the time
average will be important as the integrand will not ``really
converge'' to $S$ but it will show very rare large fluctuations which,
however, are as a rule doomed to occur at time intervals larger than
the age of the Universe, \ie do not occur {\it at all} for ``all''
purposes (I suppose that the number of particles of the system is
large, say $>10^3$): neglecting such impossible events would in fact
dispense from considering the time average in the above limit
relation. However it is not clear that there should be a universal
definition of such ``{\it Lyapunov functions}'' or ``{\it
$H$--functions}''. I think that they can certainly be defined on a case
by case basis but not necessarily in a general universal way: for
instance in fluids of higher density in the last formula one should
use the Resibois $H$--function, \Onlinecite{Re78}; and this is so, I
think, essentially because one cannot define an entropy content for
non equilibrium states: which is the quantity that would, otherwise,
naturally play the role of a universal $H$--function.

Note however that other views are possible if entropy and heat are
conceptually separated: this is discussed in Sect. V.

\*

\centerline{\bf V. Coarse graining and counting phase space 
} 
\centerline{\bf cells when volume is not conserved}
\*

Recently a quite general and universal definition has been proposed
identifying the $H$--function with the logarithm of the volume
occupied in phase space by the phase space points which are
macroscopically indistinguishable (\ie ``defining the same {\it
macrostate}''): this applies when the notion of macrostate is free of
ambiguities (or at least one can resolve them), and it is certainly an
interesting proposal which has already received support from numerical
experiments, \Onlinecite{GGL03}.  The value $S$ of this quantity could
be an interesting definition of entropy of the stationary state that
is eventually achieved by the evolution of all phase space points that
correspond to the same macrostate. Although $S$ might be unrelated to
the amounts of heat that are transferred in the transformations
between stationary states, calling it ``entropy'' would be justified
on the basis of its coincidence with entropy in the case of
equilibrium states and of its nature as a Lyapunov function for the
approach to stationarity. The question of the dependence on the notion
of macrostate that is used is still somewhat incompletely understood,
however.
\*

The problem, and the difficulty, of defining entropy can be studied
also from another viewpoint not unrelated to the latter one because it
also tries to give a precise definition of {\it coarse grained}
description of the phase space points: the attempt (whose outcome I do
not consider satisfactory for the purpose of defining entropy
of stationary states) is to identify entropy with the
number of phase space cells ``relevant'' at given external parameters,
taking advantage of a variational principle for the SRB statistics and
interpreting it as an {\it equidistribution property} in phase space.
The discussion will be performed at a level more technical than that
of the previous sections for the purpose of examining the proposal in
detail, following \Onlinecite{Ga01}.  \*

For simplicity I suppose a discrete time dynamics described by a
hyperbolic map $S$ (with dense trajectories: {\it chaotic hypothesis})
on a bounded surface $\O$ ({\it phase space}). Briefly we call the
dynamical system $(\O,S)$ an {\it Anosov map}; for a general discussion
see \Onlinecite{Ru99}. As in the continuous time case this implies
that, except for a volume zero set of initial data $x$, it will be
$\lim_{T\to\infty} T^{-1}\sum_{j=0}^{T-1} F(S^jx)=\int \m_{SRB}(dy)
F(y)$ for all continuous functions ({\it observables}) $F$ on $\O$.

The SRB distribution admits a rather simple representation which will
be interpreted here in terms of ``{\it coarse graining}'' of the phase
space. In fact given a partition of phase space into closed regions
$\PP=(P_1,\ldots,P_m)$ with pairwise disjoint interiors, consider a
point $x$ in phase space which is not in the set $\cal B$ of zero
volume consisting of the points that in their evolution fall on the
common boundary of two $P_i$'s. Then we can define the {\it history}
of $x$ on $\PP$, \ie a sequence $\V\x$ such that $S^k x\in
P_{\x_k}$. The sequence $\V\x$ must verify the property, called {\it
compatibility}, that $Q_{\x_k,\x_{k+1}}=1$ for all times $k$ where the
matrix $Q$ is defined to be $Q_{\x,\x'}=0$ unless there is an interior
point in $P_{\x}$ whose image is in the interior of $P_{\x'}$: in the
latter case $Q_{\x,\x'}=1$. The matrix $Q$ just tells us which sets
$P_{\x'}$ can be reached from points in $P_\x$ in one time step. Then
for Anosov maps one can find a partition (in fact infinitely many) of
phase space $\PP=(P_1,\ldots,P_m)$ so that,
\Onlinecite{Si68a,Si68b,Bo70a,Bo70b,Si72,BR75,Ru76,Ga00,Ga02},

\0(1) if $\V\x$ is a compatible sequence then there is a point $x$
such that $S^k x\in P_{\x_k}$, see (for instance) Ch. 9 in
\Onlinecite{Ga00}, (``{\it compatibility}''). The points $x$ outside
the exceptional set $\cal B$ (of zero volume) determine uniquely the
corresponding sequence $\V\x$.

\0(2) the diameter of the set of points $E(\x_{-\fra12T},\ldots,\x_{\fra12T})$
consisting of all points which between time $-\fra12T$ and $\fra12T$
visit, in their evolution, the sets $P_{\x_i}$ is bounded above by $c\,
e^{-\k T}$ for some $c,\k>0$ (\ie the code $\V\x\to x$ determines
$x$ ``{\it with exponential precision}'').

\0(3) there is a power $k$ of $Q$ such that $Q^k_{\x\x'}>0$ for all
$\x,\x'$ (``{\it transitivity}'').
\*

Hence points $x$ can be identified with sequences of symbols $\V\x$
verifying the compatibility property and the sequences of symbols
determine with exponential rapidity the point $x$ which they
represent. The existence of $\PP$ is nontrivial and rests on the
chaoticity of motions. \*

Given a finite string $\V\x=(\x_a,\x_{a+1},\ldots,\x_b)$ we define,
more generally, the set $E_{\V\x}$ to be the set of points $x$ in
phase space such that $S^jx\in P_{\x_j}$ for $j=a,a+1,\ldots,b$.  \*

Given a point $x$ the set of points whose history coincides with that
of $x$ for all $k>0$ large enough determines uniquely a smooth
invariant surface $\Si_s(x)$ through $x$ called the {\it stable
manifold} of $x$ and likewise the set of points whose history
coincides with that of $x$ for all $-k$ large enough determines
uniquely a smooth invariant surface $\Si_u(x)$ through $x$ called the
{\it unstable manifold} of $x$.  Points $y\in\Si_s(x)$ have a behavior
in the future very similar to that of $x$ and in fact the distance
between $S^ky$ and $S^kx$ is bounded by $C e^{-\l k}$ for $C,\l>0$
suitable. Likewise points $y\in\Si_u(x)$ have a behavior in the past
very similar to that of $x$, the distance between $S^{-k}y$ and
$S^{-k}x$ being bounded by $C e^{-\l k}$.

The expansion and contraction that take place near every point $x$ can
be captured by the matrices $\dpr S_u(x)$, $\dpr_s S(x)$ obtained by
restricting the matrix ({\it Jacobian matrix}) $\dpr S(x)$ of the
derivatives of $S$ to its action on the vectors tangent to the
unstable and stable manifolds through $x$: $S$ maps
$\Si_u(x)$,$\Si_s(x)$ to $\Si_u(Sx),\Si_s(Sx)$ and its derivative maps
the tangent vectors at $x$ into tangent vectors at $Sx$ while $\dpr
S_u(x),\dpr_s S(x)$ map tangent vectors to $\Si_u(x),\Si_s(x)$,
respectively, to corresponding vectors tangent at $Sx$.

A quantitative expression of the expansion and contraction is given by
the ``local expansion'' or ``local contraction'' exponents defined by
\begin{eqnarray}
\L_u(x)=&\log|\det (\dpr S)_u(x)|,\cr
\L_s(x)=&\log|\det (\dpr S)_s(x)|\label{1}
\end{eqnarray}
Then the SRB probablilities and the normalized volume (or ``normalized
Liouville measure'') are measures on the sets $E\subset\O$ which are
described in terms of the functions $\L_u(x),\L_s(x)$. Their
description is quite simple.

The exponential contraction in the past or in the future along the
unstable and stable manifolds, consequence of the hyperbolicity
assumption, implies that the diameter of the phase space subsets
$E_{\V\x}\defi E_{\x_{-T/2},\ldots,\x_{T/2}}$ tends to zero
exponentially fast (and uniformly in the choice the string $\V\x$) as
$T\to\infty$.

Given $T>0$ the collection of the non empty sets
$E_{\x_{-T/2},\ldots,\x_{T/2}}$ can be used to study the properties of
a restricted class of observables, namely those which have essentially
constant values on such small sets.  Let $T$ be a time such that the
size of the (nonempty) sets of the form
$E_{\x_{-T/2},\ldots,\x_{T/2}}$ is so small that the {\it few} physically
interesting observables can be viewed as constant inside each
$E_{\x_{-T/2},\ldots,\x_{T/2}}\equiv E_{\V\x}$. We define, if $Q$ is
the compatibility matrix,  \*

\0{\it Definition: ({\rm Coarse graining}) The sets of points of the form
$E_{\V\x}=E_{\x_{-T/2},\ldots,\x_{T/2}}$ will be called the elements of a
description of the microscopic states ``coarse grained to scale $\g$''
if $\g$ is the smallest linear dimension of the nonempty sets
$E_{\x_{-T/2},\ldots,\x_{T/2}}$ . The elements $E_{\V\x}$ of the
``coarse grained partition of phase space'' are labeled by a finite
string
$$\V \x\,\equiv\,\,(\x_{-T/2},\ldots,\x_{T/2})$$
with $\x_i=1,\ldots,m$ and $Q_{\x_i,\x_{i+1}}\equiv1$.}
\*

Define the {\it forward} and {\it backward} expansion and contraction rates as

$$U^{T/2}_{u,\pm}(x)=\sum_{j=0}^{\pm T/2} \L_u(S^jx),\ 
U^{T/2}_{s,\pm}(x)=\sum_{j=0}^{\pm T/2} \L_s(S^jx)$$
and select a point $x(\V\x)\in E_{\V\x}$ for each $\V\x$. Then the SRB
distribution $\m_{SRB}$ and the volume distribution $\m_L$ on the
phase space $\O$, which we suppose to have volume ${\cal W}=V(\O)$,
attribute to the {\it nonempty} sets $E_{\V\x}$ the respective
probabilities $\m,\m_L$

\begin{eqnarray}
\m(\V\x)\defi\m_{SRB}(E_{\V\x})\ {\rm and}\  \m_{L}(\V\x)\defi
\frac{V(E_{\V\x})}{\cal W}
\end{eqnarray}
if $V(E)$ denotes the Liouville volume of $E$. The distributions $\m,\m_L$ 
are shown to be defined by

\begin{eqnarray}
&\m(\V\x)\,=\,
h^T_{u,u}(\V\x)\cdot
e^{\big(-U_{u,-}^{T/2}(x(\V\x))-U^{T/2}_{u,+}(x(\V\x))\big)}\cr
&\m_{L}(\V\x)\,=\,
h^T_{s,u}(\V\x)\cdot 
e^{\big(U_{s,-}^{T/2}(x(\V\x))-U_{u,+}^{T/2}(x(\V\x))\big)}
\label{2}
\end{eqnarray}
and $h^T_{u,u}(\V\x)$, $h^T_{s,u}(\V\x)$ are suitable functions of
$\V\x$, {\it uniformly bounded as $\V\x, T$ vary}, which are mildly
dependent on $\V\x$; so that they can be regarded as constants for the
purpose of the present discussion, {\it cfr.} Ch. 9 in
\Onlinecite{Ga00},.

The (\ref{2}) shows that the Liouville volume weights {\it
asymmetrically} the past and the future while the SRB distribution
weighs them {\it symmetrically}.  One can say that the expansion and
contraction rates $\L_u,\L_s$ provide an ``energy function'' that
assigns relative probabilistic weights to the coarse grained cells via
(\ref{2}).

The latter analogy provides the motivation for the name ``{\it
thermodynamic formalism}'' that is often given to the mathematical
theory of chaotic systems, \Onlinecite{Ru78}.
\*

As mentioned above we have in mind that the sets $E_{\V\x}$ represent
macroscopic states, being small enough so that the physically
interesting observables have a constant value within them; and we
would like to think that they provide us with a model for a ``{\it
coarse grained}'' description of the microscopic states. The notion of
coarse graining given here is precise and, nevertheless, quite
flexible because it contains a free ``resolution parameter''
$\g$. Should one decide that the resolution $\g$ is not good enough
because one wants to study the system with higher accuracy then one
simply chooses a smaller $\g $ (and correspondingly a larger $T$).  \*

The phase space volume will generally contract with time: yet we want
to describe the evolution in terms of evolution of microscopic states,
with the aim of counting the microscopic states relevant for a given
stationary state of the system.

Therefore we divide phase space into parallelepipedal {\it cells} $\D$
of side size $\e\ll \g $ and try to discuss time evolution in terms of
them: we shall call such cells ``microscopic'' cells as we do not
associate them with any particular observable; they represent the
highest microscopic resolution. 

One can think of the new microscopic cells as physical realizations of
objects that arise in computer simulations: in the simulations cells
$\D$ are the ``digitally represented'' points with coordinates given
by a set of integers and the evolution $S$ is a {\it program} or {\it
code} simulating the solution of equations of motion suitable for the
model under study. The code operates {\it exactly} on the coordinates
(the deterministic round offs should be considered part of the
program).  The simulation will produce (generically) a chaotic
evolution ``for all practical purposes'', \ie if we only look at
``macroscopic observables'' which are constant on the coarse graining
scale $\g=e^{-\fra12\lis\l T}\ell_0$ of the partition $\PP^T$, where
$\ell_0$ is the phase space size and $\lis\l>0$ is the most
contractive line element exponent (which therefore fixes the scale of
the coarse graining, by the definition above).

A few words must be said about the precise meaning of ``linear
sizes'': in fact we are considering partitions of phase space into
sets ignoring that the coordinates have a physical meaning. Some of
them are momenta and others are positions hence they have different
physical dimensions. Therefore, assuming that we consider $N$
mass--$m$ particles in a gas at average kinetic energy per particle
$\frac32k_B \Th_0$ (note that $\Th_0$ will in general be different
from the temperature $\Th$ defined in Sec. III) and numerical density
$\r$, we imagine to measure such quantities in terms of units $\d q$
of length and $\d p$ of momentum fixed {\it a priori}, subject to the
constraint that they should be (following the recurrence times
estimate made by Boltzmann, p. 400 in \Onlinecite{Bo96}, see also
\Onlinecite{Ga00}) much smaller than, respectively, $\r^{-\fra13}$ and
$\sqrt{2m k_B \Th_0}$.  Then ${\cal W}=\ell_0^{6N}$ with $\ell_0$
proportional to $(\r^{-1/3}\sqrt{2m k_B \Th_0}/\d p\d q)^{1/2}$: \ie
if $d=6N$ is the phase space dimension, it is

$$\ell_0={\cal W}^{1/d},\  \ell_0=
\Big(\fra{\r^{-1/3}\sqrt{2m k_B \Th_0}}{\d p\d q}\Big)^{1/2}$$
The question we ask on general grounds is, see also \Onlinecite{Ga95}
\*

\0{\it Question: can we count the number of
ways in which the asymptotic state of the system can be realized
microscopically?}
\*

In equilibrium the (often) accepted answer is simple: the number is
$\NN_0={\cal W}/\e^d$, \ie just the number of cells (``ergodic
hypothesis''). This means that we think that motion will generate a
one cycle permutation of the $\NN_0$ cells $\D$, each of which is
therefore, representative of the equilibrium state. Average values of
macroscopic observables will be obtained simply as:

\begin{eqnarray*}
&\lim_{t\to\infty}t^{-1}\sum_{j=0}^{t-1} F(S^jx)=\cr
&=\NN_0^{-1}\sum_\D
F(\D)=\int_\O F(y)\m_L(dy)\label{3}
\end{eqnarray*}
According to Boltzmann the quantity:

\begin{eqnarray*}S_{B}\defi k_B\,\log{({\cal W}/\e^d)}\label{4}\end{eqnarray*}
is then, see \Onlinecite{Bo77}, proportional to the {\it physical
entropy} of our equilibrium system.

Can one extend the above view to stationary systems out of
equilibrium? In such systems the volume will no longer be preserved by
time evolution and, in fact, its contraction rate

\begin{eqnarray*}\h(x)=-\log |\det \dpr S(x)|\label{5}\end{eqnarray*}
not only does not vanish but, in general, will have a positive time
average $\lis\h $,
$$\lis\h =\lim_{N\to\infty}N^{-1}\sum_{j=0}^{N-1}\h(S^jx)=\int_\O
\h(y)\m_{SRB}(dy), $$
see \Onlinecite{Ru96}. If $\lis\h >0$ the volume will contract
indefinitely (hence the system will be called {\it dissipative}).  \*

{\it Out of equilibrium we may imagine that a similar kind of
``ergodicity'' holds: namely that the cells that represent the
stationary state form a subset of all the cells, on which evolution
acts as a {\rm one cycle} permutation}.
\*

If so the statistical properties of motions will be determined by the
{\it equidistribution} among such cells, which {\it therefore}
attributes probabilities $\r(\D)$ which maximize the quantity
$-\sum_\D \r(\D)\log\r(\D)$.  Hence the above counting question can be
related to a problem 
\*
{\it... which necessarily follows from Boltmann's
train of thought, {\rm[and]} has remained untouched. Consider an
irreversible process which, with fixed outside constraints, is passing
by itself from the nonstationary to the stationary state. Can we
characterize in any sense the resulting distribution of states as the
``relatively most probable distribution'', and can this be given in
terms of the minimum of a function which can be regarded as the
generalization...}, \Onlinecite{EE11}, footnote 239, p.103.  
\*

Considering realizations, and even simulations, of a dissipative
system we must recognize that {\it no representation of the evolution
as a map of the phase space cells, nor any code in simulations, can be
invertible}: it must happen (many times) that $S\D=S\D'$ with
$\D\ne\D'$. Clearly if $S\D=S\D'=\tilde\D$ we can think that both $\D$
and $\D'$ are not really different and only one of the two should be
taken as a representative of a microscopic state.

We can imagine ``pruning'' one after the other the ``unnecessary''
cells until the map $S$ becomes invertible. More formally each cell
$\D$ will have a motion that is {\it eventually} periodic and we
discard as ``{\it transients}'' all cells whose evolution is not
strictly periodic. We identify the remaining ``non transient'' cells
as the cells which ``are on the attracting set'' and which form a discrete
model of the attracting set for the motions. Correspondingly we say that a
coarse grained cell $E_{\V\x}$ is ``in the attracting set''  if it contains
non transient cells.

The question posed above becomes now a precise one: which is the number
of left over cells? It will be only a fraction of the
initial number $\NN_0$ of cells: and we can attempt to estimate it.

However we must take into account that the time $T$ is far smaller
than the time necessary to resolve the phase space cells.  Since the
length of time necessary to resolve points in phase space with
precision $\g$ is $T$ and since we assume that we shall be unable to
see cells with a precision higher than $\g$ then phase space
contraction implies that the number $\NN$ of phase space cells
necessary to describe the motions with an accuracy $\g$ is

\begin{eqnarray}\NN =\NN_0\, e^{- \,T\, \lis \h}
\label{10}\end{eqnarray}
because in time $T$ the phase space volume is reduced by a factor
about $e^{- \,T\, \lis \h}$ (here $\lis \h$ which {\it should be identified
with the infinite time average $\lis\h $ of the phase space contraction
rate $-\log |\det \dpr S(x)|$}). In fact our observations require the
phase space cells to maintain a well defined identity at least for a
time duration $T$ and they cannot be more than the r.h.s. of (\ref{10}).
This means that we imagine that if we make observations with precision
$\g$ only cells in a layer of width $\g$ around the attracting set
will be involved in the decription of the motion: cells further apart
are non recurrent (aperiodic).

We could from now on take $\g=\e$ and the analysis would be simpler:
it is however interesting and important to consider also the cases
$\g>\e$ because the extremely large size of the recurrence times on
the microscopic scales makes it not physically meningful to discuss
phenomena occurring on the recurrence time scale.

The picture must hold for all Markovian pavements $\PP$ and for all
$T$'s such that $\g=e^{-\fra12\lis\l T}\d$ $>\e$ if $\d\simeq \ell_0$
is the typical size of an element of the partition $\PP$: this induces
to choose $T$ to be of the order of $\lis
T=2\lis\l^{-1}\log\ell_0/\g$. And, as in equilibrium, once that $T$ is
so chosen and the requirement $\g\ge\e$ is fulfilled, {\it we shall think
that all cells evolve in time so that they visit all other $\NN$
cells.}

This is a kind of ``ergodicity'' assumption which is similar to the
corresponding assumption that in equilibrium all cells are actually
visited. Note that assuming that only a fraction of them is visited
(hence $\NN <\NN_0 e^{-T\lis\h}$) is also possible and, nevertheless,
it would give the {\it same statistics} as long as the fraction is
taken to be the same in each coarse grained volume, but it would give
a different cell count: if we are unable to probe phase space with
precision higher than $\g$ and $\g>\e$ we cannot distinguish the two
possibilities and taking the equal sign in (\ref{10}) can only be a
simplicity assumption which however comes with the warning that the
cell count might be somewhat ambiguous {\it unless} $\g=\e$. The same
ambiguity is also present in equilibrium cases.

Therefore the SRB distribution attributes the same weight to all cells
$\D$ {\it on the attracting set} and therefore it verifies a
variational property: namely if $C$ is a generic cell and $n(C)$ is
its probability in a statistical state of the system then $-\sum_C
n(C)\log n(C)$ attains a maximum when $n(C)=\NN^{-1}$.

It is interesting that the latter property can be identified with the
variational principle for SRB distributions, \Onlinecite{Ga95}.
Indeed from the general theory of the SRB distributions, see
\Onlinecite{Si68a,Si68b,Si72,Si77} and \Onlinecite{Ga00}, Chap. 9, we
know that the SRB distribution gives weight $e^{-U_{u,T} (\V\x)}$,
with $U_{u,T}(\V\x)=\sum_{j=-T/2}^{T/2} \log \L_u(S^j x(\V\x))$, to
the {\it collection of all cells} $\D\in E_{\V\x}$.  Therefore the
number of cells in the set $E_{\V\x}$ will be proportional to $\NN
e^{-U_{u,T} (\V\x)}$: which implies that the equal weight distribution
on the microscopic cells induces a weight $p_{\V\x}$ on the coarse
grained sets $E_{\V\x}$ which maximizes

\begin{eqnarray}P_B\defi\fra1T\sum_{\V\x} -p_{\V\x}\big(\log p_{\V\x}\ +
U_{u,T}(x(\V\x))\big)\label{10a}\end{eqnarray}
subject to $\sum_{\V\x}p_{\V\x}=1, p_{\V\x}\ge0$ and with the sums
running over the $\V\x$'s. This
is one of the characterizations of the SRB distribution ({\it
variational principle} of Ruelle, \Onlinecite{Ga00,Ga02,GBG04}).

Coming back to the cell count we take $\NN=\NN_0 e^{-\lis\h T}$, see
(\ref{10}), \ie we suppose that if we probe phase space with a
precision $\g$ then all cells closer than $\g$ to the attracting set
are visited by the motion. Call $-\l^-_i,\l^+_i$ the Lyapunov
exponents, $\l^\a_i>0$, so that $\lis\l= \max_{i,j} (\l^-_i,\l^+_j)$
and $\lis\h =\sum_i(\l^-_i-\l^+_i)$. Keeping in
mind that $\frac{\ell_0}\e=\NN_0^{\fra1d}$ and (\ref{10}) we define

\begin{eqnarray}
&S_{cells}=k_B\,\log \NN\,=\,k_B\, (\log\NN_0- 
\fra{2\lis\h}{\lis\l}\log\fra{\ell_0}\g)\,=\cr
&=\,k_B\,\big(1-\fra{2\lis\h}{d\lis\l}\big)\,\log \NN_0 
+\fra{2\lis\h}{d \lis\l}\,\log(\fra{\g}{\e})^d\, \label{11}
\end{eqnarray}
(note that $ \fra12d\,\lis\l\ge\lis\h$ is implied by the
$S_{cells}\ge0$ with $\e=\g$: this is an inequality which seems to be
satisfied in the examples studied in the literature).  The $S_{cells}$
will change if $\e/\g$ vary, as any choice (\eg $\e=\g$) of the size
of cells of the microscopic description is clearly arbitrary, {\it
and, unlike the equilibrium case when $\lis\h=0$, the change will be
nontrivial, \ie it will not simply be an additive constant independent
on the state of the system}. In fact $\lis\h/\lis \l$ is a dynamical
quantity and changing $\g$ (\ie changing the coarse graining
resolution keeping $\e$ fixed) will change $S_{cells}$ as $\D
S_{cells}=k_B d\,\fra2{d} \fra{\lis\h}{\lis\l}\log\frac{\g}{\g'}$.  \*

Given a precision $\g$ the quantity $S_{cells}$
measures how many ``non transient'' phase space cells must be used to
obtain a {\it faithful} representation of the attracting set and of
its statistical properties on scale $\g$. Here by ``faithful'' on
scale $\g$ we mean that all observables which are constant on such
scale will show the correct statistical properties, \ie that cells of
size larger than $\g$ will be visited with the correct SRB frequency.

Note that we are assuming that the system has a dense attactor (see
the above ``transitivity'' property); so
that the estimate in (\ref{11}) holds only as long as this is a
correct assumption: at high forcing the attracting set is likely (\ie
examples abound) to be no longer dense on phase space and the
number $\NN_0$ will have to be replaced by a smaller power of
$\ell_0$, affecting correspondingly the analysis leading to
(\ref{11}): we do not discuss this point here.  
\*

\0(i) Although eq. (\ref{11}) gives the cell count it does not seem to
deserve to be taken as a definition of entropy for stationary systems
out of equilibrium, not even for systems considered here, \ie simple
enough to admit an Anosov map as a model for their evolution.  It
appears as a notion distinct from what has become known as the ``Boltzmann
entropy'', \Onlinecite{Le93}, see also \Onlinecite{Ei22,Ep23}, because
it depends on the value of $\g,\e$ in a nontrivial way.  The notion is
also different from the Gibbs' entropy, to which it is equivalent only
in equilibrium: in nonequilibrium (dissipative) systems the
Gibbs' entropy can only be defined as $-\infty$ and perpetually
decreasing; because {\it in such systems one can define the rate at
which (Gibbs') entropy is ``created'' (``ceded to the thermostats'')
by the system} to be $\lis\h$, \ie to be the average phase space
contraction $\lis\h$, see \Onlinecite{An82,Ru99}.

\0(ii) We also see, from the above analysis, that the variational
principle that determines the SRB distribution can be identified with
the one that {\it leads to equal probability of the phase space
cells}, {\it cfr.} (\ref{10a}).  The SRB distribution appears to be
the equal probability distribution among the $\NN$ cells which are
``not transient'' or, as we say above, ``are on the attracting
set''. In equilibrium all cells are non transient (if ergodicity is
assumed) and the SRB distribution coincides with the Liouville
distribution.

\0(iii) If we could take $T\to\infty$ (hence, correspondingly, the
resolution $\g\to0$) then the distribution $\m$, which is uniform
inside each $E_{\V\x}$ but which attributes a total weight to $E_{\V\x}$
equal to $N(\V\x)=\m_{SRB}(E_{\V\x})\NN$, would become the exact SRB
distribution. However it seems conceptually more satisfactory,
imitating Boltzmann, to suppose that $\g$ is very small but $>0$ so
that $T$ will be large but not infinite.

\0(iv) By construction the quantity $P_B$ is a maximum as a function
of the quantities $p_{\V\x}$ when they have the value of the SRB
distribution, but it makes sense as a function defined on any
probability distribution over the microscopic cells $\D$. In
particular if the initial state is a single cell one can define
$p_{\V\x}(t)$ as the fraction of time the cell has spent in its
evolution up to time $t$ inside $E_{\V\x}$.  Therefore the {\it r.h.s.}
of (\ref{10a}) evaluated with $p_{\V\x}\equiv p_{\V\x}(t)$ tends in
the average to a maximum and it can be regarded as another instance of
an $H$--function in the sense of Section IV.

\0(v) By Pesin's formula the leading term as $T\to\infty$ of
$P_B$, (see (\ref{10a})), is $0$; hence the quantity $P_B$ which makes
sense for any probability distribution $p_{\underline\x}$ will be a
Lyapunov function for the approach of a probability distribution to
the SRB distribution but its value can hardly be taken as a definition
of a property of the stationary state. 

\0(vi) Let $\e\ll\g$. If we identify a microscopic initial state with
one of the phase space cells then we can consider the evolution of the
probability distribution which attributes initial weight $0$ to all
cells but the ones $C$ in the coarse grained region $E_{\V\x_0}$, and
equal weight to the latter ones. Then calling $p_C(t)$ the fraction of
time the cells initially in $E_{\V\x_0}$ spend in the cell $C$ we see
that the quantity $S(t)=-\sum_Cp_C(t)\log p_C(t)$ tends to $S_{cells}$
(because $p_C(t)$ tends to $\NN^{-1}$).  Therefore $S$ is a general
Lyapunov function, but it depends non trivially on arbitrary
parameters (except in the equilibrium case). The (\ref{11}) also
shows that there is a direct relation between the number of phase
space cells and the entropy production rate $\lis\h$.

\0(vii) The analysis in terms of cells is reminiscent, in fact, of the
methods employed to study Hausdorff dimension, the Hausdorff measure
and Pesin's formula in general hyperbolic systems, \Onlinecite{Yo94}.
A deeper understanding of the analysis appears to be linked to an
important question, raised by Ruelle, asking whether (and how) one
could possibly relate an entropy notion to the logarithm of the
Hausdorff measure of the attractor (mathematically the attracting set
is the closure of the attractor, and the latter is a set of smallest
Hausdorff dimension but still with SRB probability $1$). A pertinent
possibility is that the Hausdorff measure on the attractor is
absolutely continuous with respect to the SRB measure.  \*
\def\qq{{\V q}}\def\equ{}

\centerline{\bf VI. Generalities on heat transfer.}
\*

More generally one is interested in cases in which the external forces
depend also on velocities. This becomes necessary if one wants to
study the details of heat transfer processes. Heat transfer might
occurr {\it even in situations in which the only forces acting are due
to thermostats}. Therefore we extend the problems of the previous
sections to more general cases in which forces may depend on
velocities so that there is no longer an obvious distinction between
external forces and thermostat forces and we are free to call certain
forces active forces and others thermostat forces. Nevertheless in
concrete cases it will be often natural to give such attributes to
various terms into which forces can be decomposed.

In fact typically the external forces will be positional forces while
the thermostats forces will be modeled so that they can be decomposed
as sums

\begin{eqnarray}
\V\th_i(\qq,\dot\qq)=\sum_{p=1}^2
\V\th_i^{(p)}(\qq,\dot\qq)\label{12}
\end{eqnarray}
with $\th_i^{(p)}(\qq,\dot\qq)\ne\V0$ only if $\qq_i\in\L_p$  and
$\L_1,\ldots,\L_n$ are $n$ {\it disjoint} spatial regions. The regions
$\L^{(p)}$ represent the regions where the system is in contact with
the $p$-th thermostat.
\\
\def\BB{{\cal B}}\def\RR{{\cal R}}
\hbox{\kern2mm} The following model for a heat conducting and
electrically conducting gas is a good example of the above general
situations, \Onlinecite{Ga96,Ga96a}.
In a box $\BB\subset R^3$ (its dimensions are arbitrary and the shape
is drawn in the following figure), are enclosed $N$ particles with
mass $m=1$, interacting via a rather general pair potential, like a
hard core potential with a tail or via a Lennard Jones potential, and
they are subject to a constant force field ({\it electric field}) $E\V
u$ in the $x$ direction; the particles also collide with fixed
obstacles so arranged that no collisionless straight path can exist.
The boundary conditions are periodic in the horizontal direction and
reflecting in the vertical direction.

Adjacent to the box $\BB$ are located two boxes $\RR_+,\RR_-$
containing $N_+=N_-=N$ particles interacting with each other via a
hard core interaction, and with the particles in $\BB$ via a pair
interaction with suitably long range (shorter than the distance
between the two boxes), but are separated from the latter by a
reflecting wall. The sizes and the location of the three boxes can be
changed and they are fixed only for definiteness.

\vskip3mm
\eqfig{187.5pt}{97.5pt}{
\ins{9pt}{75pt}{$\scriptstyle \RR_-$}
\ins{120pt}{75pt}{$\scriptstyle \RR_+$}
\ins{-15pt}{22.5pt}{$\scriptstyle \BB$}
\ins{37.5pt}{78.75pt}{$\scriptstyle N_-$}
\ins{142.5pt}{78.75pt}{$\scriptstyle N_+$}
\ins{75pt}{37.5pt}{$\scriptstyle N$}}{fig1}{}
\vskip5mm
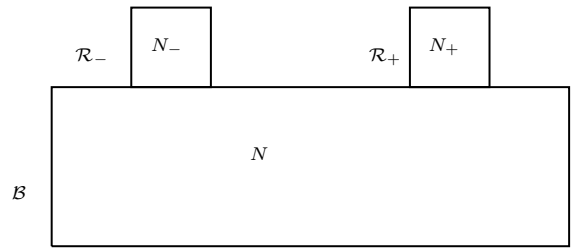
\0Fig.1: {\small The horizontal container has periodic boundary
  conditions; the two additional containers enclose the thermostatting
  sytems. On the container act an constant force establishing a
  particles flow while the temperature difference of the thermostats
  induces a heat flow producing a situation in which Onsager
  reciprocity is expected to hold in the limits $E\to0,
  \Th_+-\Th_-\to0$. \vfil}

We imagine other forces to act on the system and enforcing the
following constraints: the total kinetic energy in the "hot plate"
$\RR_+$ and in the "cold plate" $\RR_-$ are constrained to be
$\frac32Nk_B \Th_+$ and, respectively, $\frac32N_-k_B \Th_-$ where
$\Th_-$ and $\Th_+= \Th_-+\d \Th, \, \d \Th\ge0$ are the {\it
temperatures of the plates}: see Fig.1.

The equations of motion are, if $\V F_j$ are the impulsive forces due to
the hard cores plus the interparticle forces between the paricles in
the regions $\RR_\pm$ and those in $\BB$:

\begin{eqnarray}
\ddot{{\V q}}_j=&\V F_j+ E \chi_B(\V
q_j)\V u-\a_+\cr
&+\chi_{+}(\V q_j)\dot\qq_j-\a_-\chi_{-}(\V q_j)\dot\qq_j
\label{13}
\end{eqnarray}

\0where $\chi_\BB,\chi_{\pm}$ are the characteristic functions of the
regions $\BB,\RR_\pm$ and $\a_+ \,\a_-,$ are multipliers defined
so that for some preassigned $\Th_\pm$:
\begin{eqnarray}
\sum_{j=1}^{N}\chi_{\pm}(\qq_j)
\frac{\dot\qq_j^2}{2}=\frac32N k_B \Th_\pm,
\label{14}
\end{eqnarray}
\0are exact constants of motion.

\def\LL{{\cal L}} 

Let $L_E,L_+,L_-,L_+^0,L_-^0$ denote, respectively, the work per unit
time performed by the field $E$ or by the particles in the thermostats
$\RR_+,\RR_-$ on the gas in $\BB$, or by the gas in $\BB$ on the
thermostats $\RR_+,\RR_-$.

{\it We suppose that the interaction between the particles in the
thermostats $\RR_\pm$ and the ones in the main box $\BB$ are strong
enough} (compared to the field strength) so that a stationary state is
achieved and the chaotic hypothesis becomes, therefore, meaningful. In
fact it would be interesting to find a proof that one can really build
a model in which this happens: I think that the assumption should hold
if $E$ is small and the particles in $\RR_\pm$ have a sufficiently
long range interaction with the particles in the box $\RR$ and
sufficient strength; possibly even with no such extra condition in the
cases $E=0$ (pure heat conduction models). Otherwise in order to be sure
that a stationary state is reached one should add some further
constraint, \eg impose that the total energy in the main box stays
bounded: the discussion that follows would not change substantially,
as discussed in detail in \Onlinecite{Ga96}.

Let $\LL_E,\LL_+^0,\LL_-^0,\LL_+,\LL_-$ be the time averages of the
above works.

\def\QQ{{\cal Q}} 
Let $Q_\pm$ the work done per unit time by the thermostats forces in
the regions $\RR_\pm$ and $\QQ_\pm$ their time averages.  Then the
imposed conservation laws give, {\it in a stationay state},
\begin{eqnarray}
&{\LL_E+\LL_++\LL_--\LL^0_{+}-\LL^0_{-}}=0\cr
&-\QQ_++\LL^0_+-\LL_+=0,\quad
-\QQ_-+\LL^0_--\LL_-=0\label{15}\end{eqnarray}
so that one finds (by differentiating (\ref{14}) with respect to
time and by applying the equations of motion (\ref{13})) expressions
for $\a_+,\a_-$:
\begin{eqnarray}
\a_\pm=\fra{Q_\pm}{\sum \chi_\pm(\V q_j)
\dot{{\V q_j}}^2}\label{16}
\end{eqnarray}
and the phase space contraction per unit time, \ie ${k_B^{-1}}$ times the
{\it entropy production per unit time}, is:
$$\s(x)=(3N-1)(\a_++\a_-)$$
in the configuration $x$ with average (for large $N$'s)

\begin{eqnarray}
\s=\frac{\QQ_+}{k_B \Th_+}+\frac{\QQ_-}{k_B \Th_-}>0\label{17}\end{eqnarray}
Note that $\QQ_++\QQ_-=0$ if $E=0$ (hence $\LL_E=0$) and $Q_+>0, Q_-<0$.
 
The example shows that the considerations of the previous sections can
be extended to the much more general cases considered above in which
the forcing is due also to velocity dependent forces. They also give
more grounds to the identification of the entropy production rate with
the phase space contraction: see equation (\ref{17}) when $E=0$.  \*

\0{\it Remarks:} (1) It is useful to stress that in the above model
the SRB distribution is concentrated on a zero volume set in phase
space: however if one looks at the restriction of the SRB distribution
to the particles in the main box $\BB$, \ie to their statistical
properties, then it should be given by a density in the restricted
phase space: the basic mechanism for this has been discussed and
clarified in an interesting class of models in \Onlinecite{BKL03}.

\0(2) Note that according to the definitions of Sec. III if $E=0$
there is no external positional force acting on the system and
therefore the corresponding work $W$ vanishes (and $W=Q_1+Q_2=0$). So
the effective temperature is $0$: this is quite a surprising
consequence of the definitions which may mean that the definiton of
temperature proposed in Sec. III is not appropriate.  However this is
effective temperature of the combined action of the two thermostats, a
notion that is new. At the same time we see that the temperatures of
the individual thermostats are naturally identified with $\Th_+$ and
$\Th_-$ if they are defined, as in Sec. III, by the ratio between the
average work done by a thermostat force divided by its average
contribution to the phase space contraction times $k_B$.

\*
\centerline{\bf VII. Remarks.} 
\*

(1) The above analysis, if accepted, allows us to define entropy as a
notion related to heat exchanges only for the stationary states which
are in the very special class of equilibrium states. It is however
important and appropriate to mention one more study that has been made
in the attempt of defining entropy as a function of nonequilibrium
stationary states.

One can consider an evolution of a phase point under forces which up
to time $t_1$ are constant and admit a stationary SRB distribution
$\m^1$, then the forces vary between $t_1$ and $t_2$ and become again
constant after time $t_2$: during the whole process we imagine to keep
the thermostatting force varying so that the ``temperature'' of the
thermostats, as defined in Section III remains constant. If one {\it
fixed} the forces $\V E(t)$ at their value at any $t\in[t_1,t_2]$ then
the dynamics would admit a SRB distribution. Therefore we can define
for each $t$ the stationary SRB distribution $\lis\m_t$ corresponding
to the ``frozen'' forces $\V E(t)$ and, at the same time, the
(different) probability distribution $\m_t$ into which $\m^1$ evolves
in the time interval $[t^1,t]$, and we can also define $\s(\t)=\int
\s_\t(x) \m_\t( dx)$ and $\lis\s_\t=\int \s_\t(x) \lis\m_\t( dx)$.

Then a quantity which is possibly of interest is 
$$I=k_B\int_{t_1}^{t_2} (\s(t)-\lis\s_t)\, dt$$
This is a quantity that does not really depend on how long a time the
system dwells on a special value of the parameters since the integrand
will go to zero fast if the interval is too long (because $\m_t$ will
approach $\lis\m_t$). 

Does $I$ depend on the intermediate stationary states of the
transformation? Some aspects of the question have been partially
studied, \Onlinecite{Ru03}, and I interpret them as {\it suggesting}
that to first order in the variation of the force parameters (during
the intermediate times) independence might hold (I stress that this is
a conjecture). More precisely: if the variation of the forces vanishes
rapidly at $-t_1,t_2=+\infty$ (hence the evolving distribution $\m_t$
returns to $\m^1$ and the system performs a cycle) then the value of
$I$ does not depend on the actual intermediate pattern of the
variation of the forces, {\it to first order in the variations}. It is
tempting to think that the quantity $I$ could be interpreted as
entropy variation of the system in a transformation of the initial
into the final state.  However, the same work suggests that very
likely this is no longer true already to second order: going in the
direction of making difficult alternative attempts at defining entropy
variations in stationary nonequilibria on the basis of the quantity
$I$. This is a point that is likely to be clarified in the near
future. 
\*

(2) Having defined the notion of entropy production rate one can define a
``duality'' between fluxes $J_j$ and forces $E_j$ using $\s(x)$ as a
``generating function'': 
$$J_j(\V E)\,=\,k_B\,\frac{\partial \s}{\partial{E_j}}$$
which, at $\V E=\V0$, leads to Onsager's reciprocity and to
Green--Kubo's formulae for transport, \Onlinecite{Ga96,Ga96a,GR97}. 
\*

(3) We have proposed a definition of entropy production rate and of
temperature for a class of stationary states. But a new definition is
really useful if it is associated with new results: I think that such
new results may already be around and many cluster around the {\it
fluctuation theorem}. I refer, on this point, to the literature,
\Onlinecite{ECM93,GC95,BGG97,Ge98,GRS04,CG99,FN03,Ga02,Ga03}
confining myself here to recalling the theorem. This is a theorem that
holds for systems mechanically thermostatted by a thermostat which is
{\it time reversible}, \ie such that there is an isometry map $I$ of phase
space such that $I^2=$ identity and which anticommutes with the
identity, namely $I S_t=S_{-t}I$. 

For instance the model in Sect. VI is time reversible, and time
reversal is simply $I(\dot\qq,\qq)=(-\dot\qq,\qq)$. 

Suppose that $\s>0$ (see Section III) and let the average
``dimensionless'' entropy production rate over a time interval $\t$ be
$p_\t(x)=\t^{-1}\s^{-1}\int_0^\t \s(S_t x)\,dt$. Then the SRB
probability $\p_\t(p)\,dp$ that the dimensionless entropy production
rate is $p_\t(x)\in[p,p+dp]$ is related to the probability
$\p_\t(-p)\,dp$ by the relation ``FR''

$$\fra{\p_\t(p)}{\p_\t(-p)}=e^{\t\,\s\, p\,+\, O}$$
where the error term $O$ is uniformly bounded for all $\t\to\infty$,
and for all $|p|$ in a fixed closed interval contained in $[0, p^*)$,
where $1 \le p^*<\infty$ is a suitable constant. The constant $p^*$ is
a {\it non trivial dynamical quantity}, see \cite{Ga95b} where it is
defined; nevertheless it has sometimes, even recently, been confused
with the maximum value of $\s(x)/\s$, with consequent misunderstandings:
in general $\max|\s(x)|/\s> p^*$ and {\it it can be even much larger},
\Onlinecite{Ga04}.

The FR can be more clearly formulated in terms of the function
$\z(p)=\lim_{\t\to\infty}\t^{-1}\log \p_\t(p)$ and it becomes

$$\z(-p)=\z(p)-\s \,p, \qquad |p|< p^*$$
with no error term because this is a relation valid in the limit
$\t\to\infty$.
It is important to note that while $\s(x)$ is not an intrinsic
quantity, as it depends on the volume definition hence on the metric
used on phase space, the $\s$ as well as the function $\z(p)$ are
independent on the metric used and are intrinsic quantities. If one
changes the metric, say by altering it by a factor $\F(x)$, then the
variation of $\s(x)$ is $\LL \F(x)$ where $\LL$ is the Liouville
operator: hence the quantity $\t^{-1}\int_0^\t \s(S_t x)dt$ changes by
the variation $\t^{-1}(\F(S_\t x)-\F(x))$ which tends to $0$ as $\t\to\infty$.

In the latter form the FR relation can be also regarded as valid for
{\it all} $p$ if we imagine that for $|p|> p^*$ the function
$\z(p)$ is defined to be equal to $-\infty$. Indeed it is natural to
define $\z(p)=-\infty$ for $|p|$ too large, certainly for $|p|>\max_x
|\s(x)|/\s$, because such values are impossible; the value of $p^*$ above
can in fact be determined, if one follows the details in
\Onlinecite{GC95,GC95b,Ga95b}, by the condition that $\z(p)>-\infty$
for $|p|< p^*$. The value of $\z(p)$ for $p=p^*$ is a delicate matter:
one can only say that the function $\z(p)$ has to be convex, so that
if it is well defined for $p=p^*$ it can have {\it a priori} any value
between the $\lim_{p<p^*,\, p\to\p^*}\z(p)$ and $-\infty$: it is
likely that the answer is model dependent.

A way to interpret the FR for $\s=0$ is to consider it for $\V
E\ne\V0$ and to look at the limiting relations that it implies when
the forcing $\V E\to\V0$, hence $\s\to0$. As such it is very
interesting, and properly interpreted leads to Onsager reciprocity and
Green--Kubo formulae for transport coefficients in the limit of zero
forcing: hence the FR can be interpreted, in reversibly thermostatted
systems, as an extension to non zero forcing of Onsager reciprocity,
\Onlinecite{Ga96}.

As an example the fluctuation relation should hold for the models in
Sec. VI: and it would imply the usual Onsager reciprocity and
Green--Kubo relations in the limit in which $\Th_+-\Th_-, E\to0$ as
discussed in \Onlinecite{Ga96}.

One might be interested in seeing what the above relation becomes in
the case of equilibrium states or more generally in the cases in which
$\s=0$. An answer would be that it makes no sense because the
very definition of $p$ involves division by $0$.

A related question would be to ask which are the properties of the
quantity $A(x)=\t^{-1} \int_0^\t \s(S_t x)\,dt$, \ie of the
dimensional entropy creation rate: {\it but it would be uninteresting}
because one would just find that if we call $\widetilde\z(A)$ the
function analogous to $\z$ then $\widetilde\z(-A)=\widetilde\z(A)$ for
$|A|=0$ (which is of course trivially true, see also below).

\def\tende#1{\ \vtop{\ialign{##\crcr\rightarrowfill\crcr
\noalign{\kern-1pt\nointerlineskip} \hglue3.pt${\scriptstyle%
#1}$\hglue3.pt\crcr}}\,}

Nevertheless one might think that the proof of the FR in the cases
$\s>0$ could be followed (in the case of Anosov systems) with minor
adaptations to obtain some non trivial result even in the cases
$\s=0$, \Onlinecite{Ga04}.  Closer examination of this point can be
traced back to the remark that even in (Anosov) equilibrium systems
one could have a $\s(x)\not\equiv0$: in fact the value of $\s(x)$ is
identically zero in equilibrium systems {\it only if one uses the
natural metric on phase space}, whose volume element is the Liouville
volume. If one used a different metric or if one imposed some
constraint like constant kinetic energy then $\s(x)$ might be non
zero: and it will be so in general.  However in such cases the phase
space contraction would be only apparent: in fact one can show that
$\s(x)$ would have the form $\s(x)=\LL F(x)$ with $\LL$ the Liouville
operator and $F$ is a suitable function. Or, in the case of discrete
time evolution $S$, the phase space contraction $\s(x)$ would take the
form $\s(x)=F(Sx)-F(x)$, see proposition 6.4.3 in
\Onlinecite{GBG04}. So that $A(x)=\t^{-1}\int_0^\t
A(S_tx)\,d\t\,=\,\t^{-1}(F(S_\t x)-F(x))\tende{\t\to\infty}0$ and the
FR would become trivial because the distribution of $A$ becomes a
delta funtion at the origin. If one follows the proof of the FR,
\Onlinecite{Ga95b}, under the additional assumption $\s=0$ and with
the appropriate modifications one ends up, {\it if mistakes are
avoided}, \Onlinecite{Ga04}, with a trivial statement that FR holds
for $p=0$ only.

Forgetting or misinterpreting the restrictions of validity for the FR
(\ie $p<p^*$ or $|A|<p^*\s$) as well as neglecting the problems
arising in its intepretation for the distribution of $A$ in the cases
$\s=0$ has led to a variety of mistakes, paradoxes and logical errors,
\Onlinecite{Ga04}. In fact in applications taking the restrictions into
account is essential, see for instance \Onlinecite{ZRA03}.


\vglue3mm 
\0{\it Acknowledgements: The above review has been stimulated
by discussions held mostly at Rutgers University in the course of the
last few years: involving F.Bonet\-to, O. Costin, S.Gold\-stein,
J.Lebowitz, D. Ruelle, E. Speer, F. Zamponi. I am particularly
indebted to A. Giuliani for his comments, corrections and suggestions
and to E.G.D. Cohen for the ideas discussed in the three introductory
sections (an expanded version of {\rm\Onlinecite{GC04}}).\hfill}

\kern3mm


\begin{thebibliography}{60}
\expandafter\ifx\csname natexlab\endcsname\relax\def\natexlab#1{#1}\fi
\expandafter\ifx\csname bibnamefont\endcsname\relax
  \def\bibnamefont#1{#1}\fi
\expandafter\ifx\csname bibfnamefont\endcsname\relax
  \def\bibfnamefont#1{#1}\fi
\expandafter\ifx\csname citenamefont\endcsname\relax
  \def\citenamefont#1{#1}\fi
\expandafter\ifx\csname url\endcsname\relax
  \def\url#1{\texttt{#1}}\fi
\expandafter\ifx\csname urlprefix\endcsname\relax\def\urlprefix{URL }\fi
\providecommand{\bibinfo}[2]{#2}
\providecommand{\eprint}[2][]{\url{#2}}

\bibitem[{\citenamefont{Zemansky}(1957)}]{Ze68}
\bibinfo{author}{\bibfnamefont{M.}~\bibnamefont{Zemansky}},
  \emph{\bibinfo{title}{Heat and thermodynamics}}
  (\bibinfo{publisher}{McGraw-Hill}, \bibinfo{address}{New-York},
  \bibinfo{year}{1957}).

\bibitem[{\citenamefont{Gallavotti}(2004{\natexlab{a}})}]{Ga03}
\bibinfo{author}{\bibfnamefont{G.}~\bibnamefont{Gallavotti}},
  \emph{\bibinfo{title}{Nonequilibrium Thermodynamics ? Twentyseven comments}},
  Meteorological and geophysical fluid dynamics, ed. {W}. {S}chr{\"o}der
  (\bibinfo{publisher}{Science}, \bibinfo{year}{2004}{\natexlab{a}}).

\bibitem[{\citenamefont{Gallavotti}(2002)}]{Ga02}
\bibinfo{author}{\bibfnamefont{G.}~\bibnamefont{Gallavotti}},
  \emph{\bibinfo{title}{Foundations of Fluid Dynamics}}
  (\bibinfo{publisher}{Springer Verlag}, \bibinfo{address}{Berlin},
  \bibinfo{year}{2002}).

\bibitem[{\citenamefont{de~Groot and Mazur}(1984)}]{DGM84}
\bibinfo{author}{\bibfnamefont{S.}~\bibnamefont{de~Groot}} \bibnamefont{and}
  \bibinfo{author}{\bibfnamefont{P.}~\bibnamefont{Mazur}},
  \emph{\bibinfo{title}{Non equilibrium thermodynamics}}
  (\bibinfo{publisher}{Dover}, \bibinfo{address}{Mineola, NY},
  \bibinfo{year}{1984}).

\bibitem[{\citenamefont{Gallavotti and Cohen}(1995{\natexlab{a}})}]{GC95}
\bibinfo{author}{\bibfnamefont{G.}~\bibnamefont{Gallavotti}} \bibnamefont{and}
  \bibinfo{author}{\bibfnamefont{E.}~\bibnamefont{Cohen}},
  \bibinfo{journal}{Physical Review Letters} \textbf{\bibinfo{volume}{74}},
  \bibinfo{pages}{2694} (\bibinfo{year}{1995}{\natexlab{a}}).

\bibitem[{\citenamefont{Gallavotti}(2000)}]{Ga00}
\bibinfo{author}{\bibfnamefont{G.}~\bibnamefont{Gallavotti}},
  \emph{\bibinfo{title}{Statistical Mechanics. A short treati\-se}}
  (\bibinfo{publisher}{Springer Verlag}, \bibinfo{address}{Berlin},
  \bibinfo{year}{2000}).

\bibitem[{\citenamefont{Becker}(1964)}]{Be64}
\bibinfo{author}{\bibfnamefont{R.}~\bibnamefont{Becker}},
  \emph{\bibinfo{title}{Electromagnetic fields and interactions}}
  (\bibinfo{publisher}{Blaisdell}, \bibinfo{address}{New-York},
  \bibinfo{year}{1964}).

\bibitem[{\citenamefont{Evans and Morriss}(1990)}]{EM90}
\bibinfo{author}{\bibfnamefont{D.}~\bibnamefont{Evans}} \bibnamefont{and}
  \bibinfo{author}{\bibfnamefont{G.}~\bibnamefont{Morriss}},
  \emph{\bibinfo{title}{Statistical Mechanics of Nonequilibrium Fluids}}
  (\bibinfo{publisher}{Academic Press}, \bibinfo{address}{New-York},
  \bibinfo{year}{1990}).

\bibitem[{\citenamefont{Kurchan}(1998)}]{Ku98}
\bibinfo{author}{\bibfnamefont{J.}~\bibnamefont{Kurchan}},
  \bibinfo{journal}{Journal of Physics A} \textbf{\bibinfo{volume}{31}},
  \bibinfo{pages}{3719} (\bibinfo{year}{1998}).

\bibitem[{\citenamefont{Lebowitz and Spohn}(1999)}]{LS99}
\bibinfo{author}{\bibfnamefont{J.}~\bibnamefont{Lebowitz}} \bibnamefont{and}
  \bibinfo{author}{\bibfnamefont{H.}~\bibnamefont{Spohn}},
  \bibinfo{journal}{Journal of Statistical Physics}
  \textbf{\bibinfo{volume}{95}}, \bibinfo{pages}{333} (\bibinfo{year}{1999}).

\bibitem[{\citenamefont{Maes}(1999)}]{Ma99}
\bibinfo{author}{\bibfnamefont{C.}~\bibnamefont{Maes}},
  \bibinfo{journal}{Journal of Statistical Physics}
  \textbf{\bibinfo{volume}{95}}, \bibinfo{pages}{367} (\bibinfo{year}{1999}).

\bibitem[{\citenamefont{Depken}(2003)}]{De02}
\bibinfo{author}{\bibfnamefont{M.}~\bibnamefont{Depken}},
  \bibinfo{journal}{cond-mat/0209284} pp. \bibinfo{pages}{1--10}
  (\bibinfo{year}{2003}).

\bibitem[{\citenamefont{Ruelle}(1995)}]{Ru95}
\bibinfo{author}{\bibfnamefont{D.}~\bibnamefont{Ruelle}},
  \emph{\bibinfo{title}{Turbulence, strange attractors and chaos}}
  (\bibinfo{publisher}{World Scientific}, \bibinfo{address}{New-York},
  \bibinfo{year}{1995}).

\bibitem[{\citenamefont{Gallavotti
  et~al.}(2004{\natexlab{a}})\citenamefont{Gallavotti, Bonetto, and
  Gentile}}]{GBG04}
\bibinfo{author}{\bibfnamefont{G.}~\bibnamefont{Gallavotti}},
  \bibinfo{author}{\bibfnamefont{F.}~\bibnamefont{Bonetto}}, \bibnamefont{and}
  \bibinfo{author}{\bibfnamefont{G.}~\bibnamefont{Gentile}},
  \emph{\bibinfo{title}{Aspects of the ergodic, qualitative and statistical
  theory of motion}} (\bibinfo{publisher}{Springer Verlag},
  \bibinfo{address}{Berlin}, \bibinfo{year}{2004}{\natexlab{a}}).

\bibitem[{\citenamefont{Boltzmann}(1968{\natexlab{a}})}]{Bo84}
\bibinfo{author}{\bibfnamefont{L.}~\bibnamefont{Boltzmann}},
  \emph{\bibinfo{title}{{\"U}ber die {E}igenshaften monozyklischer und anderer
  damit verwandter {S}ysteme}}, vol.~\bibinfo{volume}{3} of
  \emph{\bibinfo{series}{{W}issenschafltliche {A}bhandlungen}}
  (\bibinfo{publisher}{Chelsea}, \bibinfo{address}{New-York},
  \bibinfo{year}{1968}{\natexlab{a}}).

\bibitem[{\citenamefont{Brush}(2003)}]{Br03}
\bibinfo{author}{\bibfnamefont{S.}~\bibnamefont{Brush}},
  \emph{\bibinfo{title}{History of modern physical sciences: The kinetic theory
  of gases}} (\bibinfo{publisher}{Imperial College Press},
  \bibinfo{address}{London}, \bibinfo{year}{2003}).

\bibitem[{\citenamefont{Gregory}(2003)}]{Gr798}
\bibinfo{author}{\bibfnamefont{G.}~\bibnamefont{Gregory}},
  \emph{\bibinfo{title}{The existence of fire}}, vol.~\bibinfo{volume}{1} of
  \emph{\bibinfo{series}{History of modern physical sciences: The kinetic
  theory of gases, ed. {S. B}rush}} (\bibinfo{publisher}{Imperial College
  Press}, \bibinfo{address}{London}, \bibinfo{year}{2003}).

\bibitem[{\citenamefont{Andrej}(1982)}]{An82}
\bibinfo{author}{\bibfnamefont{L.}~\bibnamefont{Andrej}},
  \bibinfo{journal}{Physics Letters} \textbf{\bibinfo{volume}{111A}},
  \bibinfo{pages}{45} (\bibinfo{year}{1982}).

\bibitem[{\citenamefont{Ruelle}(1999)}]{Ru99}
\bibinfo{author}{\bibfnamefont{D.}~\bibnamefont{Ruelle}},
  \bibinfo{journal}{Journal of Statistical Physics}
  \textbf{\bibinfo{volume}{95}}, \bibinfo{pages}{393} (\bibinfo{year}{1999}).

\bibitem[{\citenamefont{Ruelle}(1996)}]{Ru96}
\bibinfo{author}{\bibfnamefont{D.}~\bibnamefont{Ruelle}},
  \bibinfo{journal}{Journal of Statistical Physics}
  \textbf{\bibinfo{volume}{85}}, \bibinfo{pages}{1} (\bibinfo{year}{1996}).

\bibitem[{\citenamefont{Feitosa and Menon}(2003)}]{FN03}
\bibinfo{author}{\bibfnamefont{K.}~\bibnamefont{Feitosa}} \bibnamefont{and}
  \bibinfo{author}{\bibfnamefont{N.}~\bibnamefont{Menon}},
  \bibinfo{journal}{cond-mat/0308212}  (\bibinfo{year}{2003}).

\bibitem[{\citenamefont{Cugliandolo et~al.}(1997)\citenamefont{Cugliandolo,
  Kurchan, and L.Peliti}}]{CKP97}
\bibinfo{author}{\bibfnamefont{L.}~\bibnamefont{Cugliandolo}},
  \bibinfo{author}{\bibfnamefont{J.}~\bibnamefont{Kurchan}}, \bibnamefont{and}
  \bibinfo{author}{\bibnamefont{L.Peliti}}, \bibinfo{journal}{Physical Review
  E} \textbf{\bibinfo{volume}{55}}, \bibinfo{pages}{3898}
  (\bibinfo{year}{1997}).

\bibitem[{\citenamefont{Cugliandolo}(2002)}]{Cu02}
\bibinfo{author}{\bibfnamefont{L.}~\bibnamefont{Cugliandolo}},
  \bibinfo{journal}{cond-mat/0210312}  (\bibinfo{year}{2002}).

\bibitem[{\citenamefont{Gallavotti}(1996{\natexlab{a}})}]{Ga96}
\bibinfo{author}{\bibfnamefont{G.}~\bibnamefont{Gallavotti}},
  \bibinfo{journal}{Journal of Statistical Physics}
  \textbf{\bibinfo{volume}{84}}, \bibinfo{pages}{899}
  (\bibinfo{year}{1996}{\natexlab{a}}).

\bibitem[{\citenamefont{Gallavotti}(1996{\natexlab{b}})}]{Ga96a}
\bibinfo{author}{\bibfnamefont{G.}~\bibnamefont{Gallavotti}},
  \bibinfo{journal}{Physical Review Letters} \textbf{\bibinfo{volume}{77}},
  \bibinfo{pages}{4334} (\bibinfo{year}{1996}{\natexlab{b}}).

\bibitem[{\citenamefont{Gallavotti and Ruelle}(1997)}]{GR97}
\bibinfo{author}{\bibfnamefont{G.}~\bibnamefont{Gallavotti}} \bibnamefont{and}
  \bibinfo{author}{\bibfnamefont{D.}~\bibnamefont{Ruelle}},
  \bibinfo{journal}{Com\-mu\-ni\-ca\-tions in Mathematical Physics}
  \textbf{\bibinfo{volume}{190}}, \bibinfo{pages}{279} (\bibinfo{year}{1997}).

\bibitem[{\citenamefont{Zamponi et~al.}(2004)\citenamefont{Zamponi, Ruocco, and
  An\-gelani}}]{ZRA04}
\bibinfo{author}{\bibfnamefont{F.}~\bibnamefont{Zamponi}},
  \bibinfo{author}{\bibfnamefont{G.}~\bibnamefont{Ruocco}}, \bibnamefont{and}
  \bibinfo{author}{\bibfnamefont{L.}~\bibnamefont{An\-gelani}},
  \bibinfo{journal}{cond-mat/0403579}  (\bibinfo{year}{2004}).

\bibitem[{\citenamefont{Lebowitz}(1993)}]{Le93}
\bibinfo{author}{\bibfnamefont{J.}~\bibnamefont{Lebowitz}},
  \bibinfo{journal}{Physics Today} \textbf{\bibinfo{volume}{September}},
  \bibinfo{pages}{32} (\bibinfo{year}{1993}).

\bibitem[{\citenamefont{Goldstein and Lebowitz}(2003)}]{GL03}
\bibinfo{author}{\bibfnamefont{S.}~\bibnamefont{Goldstein}} \bibnamefont{and}
  \bibinfo{author}{\bibfnamefont{J.}~\bibnamefont{Lebowitz}},
  \bibinfo{journal}{cond-math/0306078}  (\bibinfo{year}{2003}).

\bibitem[{\citenamefont{Garrido et~al.}(2003)\citenamefont{Garrido, Goldstein,
  and Lebowitz}}]{GGL03}
\bibinfo{author}{\bibfnamefont{P.}~\bibnamefont{Garrido}},
  \bibinfo{author}{\bibfnamefont{S.}~\bibnamefont{Goldstein}},
  \bibnamefont{and} \bibinfo{author}{\bibfnamefont{J.}~\bibnamefont{Lebowitz}},
  \bibinfo{journal}{cond-math/0310575}  (\bibinfo{year}{2003}).

\bibitem[{\citenamefont{Resibois}(1978)}]{Re78}
\bibinfo{author}{\bibfnamefont{P.}~\bibnamefont{Resibois}},
  \bibinfo{journal}{Journal of Statistical Physics}
  \textbf{\bibinfo{volume}{19}}, \bibinfo{pages}{593} (\bibinfo{year}{1978}).

\bibitem[{\citenamefont{Gallavotti}(2001)}]{Ga01}
\bibinfo{author}{\bibfnamefont{G.}~\bibnamefont{Gallavotti}},
  \bibinfo{journal}{Communication in Mathematical Physics}
  \textbf{\bibinfo{volume}{224}}, \bibinfo{pages}{107} (\bibinfo{year}{2001}).

\bibitem[{\citenamefont{Sinai}(1968{\natexlab{a}})}]{Si68a}
\bibinfo{author}{\bibfnamefont{Y.}~\bibnamefont{Sinai}},
  \bibinfo{journal}{Functional Analysis and Applications}
  \textbf{\bibinfo{volume}{2}}, \bibinfo{pages}{64}
  (\bibinfo{year}{1968}{\natexlab{a}}).

\bibitem[{\citenamefont{Sinai}(1968{\natexlab{b}})}]{Si68b}
\bibinfo{author}{\bibfnamefont{Y.}~\bibnamefont{Sinai}},
  \bibinfo{journal}{Functional analysis and Applications}
  \textbf{\bibinfo{volume}{2}}, \bibinfo{pages}{70}
  (\bibinfo{year}{1968}{\natexlab{b}}).

\bibitem[{\citenamefont{Bowen}(1970{\natexlab{a}})}]{Bo70a}
\bibinfo{author}{\bibfnamefont{R.}~\bibnamefont{Bowen}},
  \bibinfo{journal}{American Journal of Mathematics}
  \textbf{\bibinfo{volume}{92}}, \bibinfo{pages}{725}
  (\bibinfo{year}{1970}{\natexlab{a}}).

\bibitem[{\citenamefont{Bowen}(1970{\natexlab{b}})}]{Bo70b}
\bibinfo{author}{\bibfnamefont{R.}~\bibnamefont{Bowen}},
  \bibinfo{journal}{American Journal of Mathematics}
  \textbf{\bibinfo{volume}{92}}, \bibinfo{pages}{907}
  (\bibinfo{year}{1970}{\natexlab{b}}).

\bibitem[{\citenamefont{Sinai}(1972)}]{Si72}
\bibinfo{author}{\bibfnamefont{Y.}~\bibnamefont{Sinai}},
  \bibinfo{journal}{Russian Mathematical Surveys}
  \textbf{\bibinfo{volume}{27}}, \bibinfo{pages}{21} (\bibinfo{year}{1972}).

\bibitem[{\citenamefont{Bowen and Ruelle}(1975)}]{BR75}
\bibinfo{author}{\bibfnamefont{R.}~\bibnamefont{Bowen}} \bibnamefont{and}
  \bibinfo{author}{\bibfnamefont{D.}~\bibnamefont{Ruelle}},
  \bibinfo{journal}{Inventiones Mathematicae} \textbf{\bibinfo{volume}{29}},
  \bibinfo{pages}{181} (\bibinfo{year}{1975}).

\bibitem[{\citenamefont{Ruelle}(1976)}]{Ru76}
\bibinfo{author}{\bibfnamefont{D.}~\bibnamefont{Ruelle}},
  \bibinfo{journal}{American Journal of Mathematics}
  \textbf{\bibinfo{volume}{98}}, \bibinfo{pages}{619} (\bibinfo{year}{1976}).

\bibitem[{\citenamefont{Ruelle}(1978)}]{Ru78}
\bibinfo{author}{\bibfnamefont{D.}~\bibnamefont{Ruelle}},
  \emph{\bibinfo{title}{Thermodynamic formalism}} (\bibinfo{publisher}{Addison
  Wesley}, \bibinfo{address}{Reading}, \bibinfo{year}{1978}).

\bibitem[{\citenamefont{Boltzmann}(2003)}]{Bo96}
\bibinfo{author}{\bibfnamefont{L.}~\bibnamefont{Boltzmann}},
  \emph{\bibinfo{title}{Reply to Zermelo's Remarks on the theory of heat}},
  vol.~\bibinfo{volume}{1} of \emph{\bibinfo{series}{History of modern physical
  sciences: The kinetic theory of gases, ed. {S. B}rush}}
  (\bibinfo{publisher}{Imperial College Press}, \bibinfo{address}{London},
  \bibinfo{year}{2003}).

\bibitem[{\citenamefont{Gallavotti}(1995{\natexlab{a}})}]{Ga95}
\bibinfo{author}{\bibfnamefont{G.}~\bibnamefont{Gallavotti}},
  \bibinfo{journal}{Journal of Statistical Physics}
  \textbf{\bibinfo{volume}{78}}, \bibinfo{pages}{1571}
  (\bibinfo{year}{1995}{\natexlab{a}}).

\bibitem[{\citenamefont{Boltzmann}(1968{\natexlab{b}})}]{Bo77}
\bibinfo{author}{\bibfnamefont{L.}~\bibnamefont{Boltzmann}},
  \emph{\bibinfo{title}{{\"U}ber die {B}eziehung zwischen dem zwei\-ten
  {H}aupt\-satze der mechanischen {W}{\"a}rmetheo\-rie und der
  {W}ahrscheinlichkeitsrechnung, respektive den {S}{\"a}tz\-en {\"u}ber das
  {W}{\"a}rme\-gleichgewicht}}, vol.~\bibinfo{volume}{2} of
  \emph{\bibinfo{series}{{W}is\-sen\-schaft\-li\-che {A}bhandlungen, ed. {F}.
  {H}asen{\"o}hrl}} (\bibinfo{publisher}{Chelsea}, \bibinfo{address}{New York},
  \bibinfo{year}{1968}{\natexlab{b}}).

\bibitem[{\citenamefont{Ehrenfest and Ehrenfest}(1990)}]{EE11}
\bibinfo{author}{\bibfnamefont{P.}~\bibnamefont{Ehrenfest}} \bibnamefont{and}
  \bibinfo{author}{\bibfnamefont{T.}~\bibnamefont{Ehrenfest}},
  \emph{\bibinfo{title}{The conceptual foundations of the statistical approach
  in Mechanics}} (\bibinfo{publisher}{Dover}, \bibinfo{address}{New York},
  \bibinfo{year}{1990}).

\bibitem[{\citenamefont{Sinai}(1977)}]{Si77}
\bibinfo{author}{\bibfnamefont{Y.}~\bibnamefont{Sinai}},
  \emph{\bibinfo{title}{Lectures in ergodic theory}}, Lecture notes in
  Mathematics (\bibinfo{publisher}{Princeton University Press},
  \bibinfo{address}{Princeton}, \bibinfo{year}{1977}).

\bibitem[{\citenamefont{Einstein}(1922)}]{Ei22}
\bibinfo{author}{\bibfnamefont{E.}~\bibnamefont{Einstein}},
  \bibinfo{journal}{Annalen der Physik} \textbf{\bibinfo{volume}{69}},
  \bibinfo{pages}{241} (\bibinfo{year}{1922}).

\bibitem[{\citenamefont{Epstein}(1923)}]{Ep23}
\bibinfo{author}{\bibfnamefont{P.}~\bibnamefont{Epstein}},
  \bibinfo{journal}{Zeitschrift f$\ddot u$r Physik}
  \textbf{\bibinfo{volume}{54}}, \bibinfo{pages}{710} (\bibinfo{year}{1923}).

\bibitem[{\citenamefont{Young}(1995)}]{Yo94}
\bibinfo{author}{\bibfnamefont{L.}~\bibnamefont{Young}}, \bibinfo{journal}{in
  Real and complex dynamical systems, ed. B. Branner, P. Hjorth}
  (\bibinfo{year}{1995}).

\bibitem[{\citenamefont{Bonetto et~al.}(2002)\citenamefont{Bonetto, Kupiainen,
  and Lebowitz}}]{BKL03}
\bibinfo{author}{\bibfnamefont{F.}~\bibnamefont{Bonetto}},
  \bibinfo{author}{\bibfnamefont{A.}~\bibnamefont{Kupiainen}},
  \bibnamefont{and} \bibinfo{author}{\bibfnamefont{J.}~\bibnamefont{Lebowitz}},
  \bibinfo{journal}{Preprint}  (\bibinfo{year}{2002}).

\bibitem[{\citenamefont{Ruelle}(2003)}]{Ru03}
\bibinfo{author}{\bibfnamefont{D.}~\bibnamefont{Ruelle}},
  \bibinfo{journal}{Proceedings of the National Academy of Sciences}
  \textbf{\bibinfo{volume}{100}}, \bibinfo{pages}{30054}
  (\bibinfo{year}{2003}).

\bibitem[{\citenamefont{Evans et~al.}(1993)\citenamefont{Evans, Cohen, and
  Morriss}}]{ECM93}
\bibinfo{author}{\bibfnamefont{D.}~\bibnamefont{Evans}},
  \bibinfo{author}{\bibfnamefont{E.}~\bibnamefont{Cohen}}, \bibnamefont{and}
  \bibinfo{author}{\bibfnamefont{G.}~\bibnamefont{Morriss}},
  \bibinfo{journal}{Physical Review Letters} \textbf{\bibinfo{volume}{70}},
  \bibinfo{pages}{2401} (\bibinfo{year}{1993}).

\bibitem[{\citenamefont{Bonetto et~al.}(1997)\citenamefont{Bonetto, Gallavotti,
  and Garrido}}]{BGG97}
\bibinfo{author}{\bibfnamefont{F.}~\bibnamefont{Bonetto}},
  \bibinfo{author}{\bibfnamefont{G.}~\bibnamefont{Gallavotti}},
  \bibnamefont{and} \bibinfo{author}{\bibfnamefont{P.}~\bibnamefont{Garrido}},
  \bibinfo{journal}{Physica D} \textbf{\bibinfo{volume}{105}},
  \bibinfo{pages}{226} (\bibinfo{year}{1997}).

\bibitem[{\citenamefont{Gentile}(1998)}]{Ge98}
\bibinfo{author}{\bibfnamefont{G.}~\bibnamefont{Gentile}},
  \bibinfo{journal}{Forum Mathematicum} \textbf{\bibinfo{volume}{10}},
  \bibinfo{pages}{89} (\bibinfo{year}{1998}).

\bibitem[{\citenamefont{Gallavotti
  et~al.}(2004{\natexlab{b}})\citenamefont{Gallavotti, Rondoni, and
  Segre}}]{GRS04}
\bibinfo{author}{\bibfnamefont{G.}~\bibnamefont{Gallavotti}},
  \bibinfo{author}{\bibfnamefont{L.}~\bibnamefont{Rondoni}}, \bibnamefont{and}
  \bibinfo{author}{\bibfnamefont{E.}~\bibnamefont{Segre}},
  \bibinfo{journal}{Physica D} \textbf{\bibinfo{volume}{187}},
  \bibinfo{pages}{358} (\bibinfo{year}{2004}{\natexlab{b}}).

\bibitem[{\citenamefont{Cohen and G.Gallavotti}(1999)}]{CG99}
\bibinfo{author}{\bibfnamefont{E.}~\bibnamefont{Cohen}} \bibnamefont{and}
  \bibinfo{author}{\bibnamefont{G.Gallavotti}}, \bibinfo{journal}{Journal of
  Statistical Physics} \textbf{\bibinfo{volume}{96}}, \bibinfo{pages}{1343}
  (\bibinfo{year}{1999}).

\bibitem[{\citenamefont{Gallavotti}(1995{\natexlab{b}})}]{Ga95b}
\bibinfo{author}{\bibfnamefont{G.}~\bibnamefont{Gallavotti}},
  \bibinfo{journal}{Mathematical Physics Electronic Journal (MPEJ)}
  \textbf{\bibinfo{volume}{1}}, \bibinfo{pages}{1}
  (\bibinfo{year}{1995}{\natexlab{b}}).

\bibitem[{\citenamefont{Gallavotti}(2004{\natexlab{b}})}]{Ga04}
\bibinfo{author}{\bibfnamefont{G.}~\bibnamefont{Gallavotti}},
  \bibinfo{journal}{cond-mat 0402676}  (\bibinfo{year}{2004}{\natexlab{b}}).

\bibitem[{\citenamefont{Gallavotti and Cohen}(1995{\natexlab{b}})}]{GC95b}
\bibinfo{author}{\bibfnamefont{G.}~\bibnamefont{Gallavotti}} \bibnamefont{and}
  \bibinfo{author}{\bibfnamefont{E.}~\bibnamefont{Cohen}},
  \bibinfo{journal}{Journal of Statistical Physics}
  \textbf{\bibinfo{volume}{80}}, \bibinfo{pages}{931}
  (\bibinfo{year}{1995}{\natexlab{b}}).

\bibitem[{\citenamefont{Zamponi et~al.}(2003)\citenamefont{Zamponi, Ruocco, and
  An\-gelani}}]{ZRA03}
\bibinfo{author}{\bibfnamefont{F.}~\bibnamefont{Zamponi}},
  \bibinfo{author}{\bibfnamefont{G.}~\bibnamefont{Ruocco}}, \bibnamefont{and}
  \bibinfo{author}{\bibfnamefont{L.}~\bibnamefont{An\-gelani}},
  \bibinfo{journal}{cond-mat/0311583}  (\bibinfo{year}{2003}).

\bibitem[{\citenamefont{Gallavotti and Cohen}(2004)}]{GC04}
\bibinfo{author}{\bibfnamefont{G.}~\bibnamefont{Gallavotti}} \bibnamefont{and}
  \bibinfo{author}{\bibfnamefont{E.}~\bibnamefont{Cohen}},
  \bibinfo{journal}{Physical Review E} \textbf{\bibinfo{volume}{69}},
  \bibinfo{pages}{035104 (+4)} (\bibinfo{year}{2004}).

\end{thebibliography}

\bibliographystyle{apsrev}

\def\revtexz{{\bf
R\lower1mm\hbox{E}V\lower1mm\hbox{T}E\lower1mm\hbox{X}}} 
\*

\0e-mail: {\tt giovanni.gallavotti@roma1.infn.it}\\
web: {\tt http://ipparco.roma1.infn.it}\\
Piscataway, NJ, November 2003

\0\revtex
\end{document}